\documentclass[sigplan, screen, nonacm, 10pt]{acmart}

\usepackage{times,color,graphicx,amsmath,array, subcaption}

\usepackage{times,helvet,courier,xspace}
\usepackage{boxedminipage,multirow,endnotes,balance}

\usepackage{pdfpages}
\usepackage{rotating}
\usepackage{colortbl}
\usepackage{transparent}
\usepackage{appendix}
\usepackage{listings}

\newcommand{\squishlist}{
   \begin{list}{$\bullet$}
    { \setlength{\itemsep}{0pt}      \setlength{\parsep}{3pt}
      \setlength{\topsep}{3pt}       \setlength{\partopsep}{0pt}
      \setlength{\leftmargin}{1em} \setlength{\labelwidth}{1em}
      \setlength{\labelsep}{0.5em} } }
\newcommand{\squishend}{
    \end{list}  }
\newcommand{\name}{Revelio}

\definecolor{editorGreen}{rgb}{0, 0, 0.5}
\definecolor{editorRed}{rgb}{0.5, 0, 0}
\usepackage{ifthen}

  {\begin{list}{$\bullet$}%
     {\setlength{\parsep}{0pt}%
      \setlength{\topsep}{0pt}%
      \setlength{\itemsep}{0pt}}}%
  {\end{list}}

\makeatletter

\global\def\section{\@startsection {section}{1}{\z@}%
                                   {-1.5ex \@plus -0.8ex \@minus -.1ex}%
                                   {0.6ex \@plus.2ex}
                                   {\normalfont\bfseries\scshape\fontsize{11}{13}\selectfont}}
\global\def\subsection{\@startsection{subsection}{2}{\z@}%
                                     {-1.25ex\@plus -0.8ex \@minus -.1ex}%
                                     {0.3ex \@plus .1ex}
                                     {\normalfont\bfseries\fontsize{10}{12}\selectfont}}
\global\def\subsubsection{\@startsection{subsubsection}{3}{\z@}%
                                     {-1ex\@plus -1ex \@minus -.1ex}%
                                     {0.1ex \@plus .1ex}
                                     {\normalfont\itshape\fontsize{10}{12}\selectfont}}

\usepackage{bibspacing}
\newcommand{\tightcaption}[1]{\vspace{-11pt}\caption{{\bf \small #1}}
\vspace{-15pt}
}
\newcommand{\para}[1]{\smallskip\noindent{\bf #1}}

\newcommand{\Prob}{{\mathbf P}}

\author{Pradeep Dogga}
\affiliation{%
 \institution{UCLA}
}

\author{Karthik Narasimhan}
\affiliation{%
 \institution{Princeton University}
}

\author{Anirudh Sivaraman}
\affiliation{%
 \institution{NYU}
}

\author{Shiv Kumar Saini}
\affiliation{%
 \institution{Adobe Research, India}
}

\author{George Varghese}
\affiliation{%
 \institution{UCLA}
}

\author{Ravi Netravali}
\affiliation{%
 \institution{Princeton University}
}

\def\compactify{\itemsep=0pt \topsep=0pt \partopsep=0pt \parsep=0pt \leftmargin=17pt}
\let\latexusecounter=\usecounter

\newenvironment{CompactEnumerate}
  {\def\usecounter{\compactify\latexusecounter}
   \begin{enumerate}}
  {\end{enumerate}\let\usecounter=\latexusecounter}
\newenvironment{parafont}{\fontfamily{ptm}\selectfont}{}
\newcommand{\Para}[1]{\vspace{4pt}\noindent\begin{parafont}\textbf{\textit{#1}}\end{parafont}}

\begin{document}
\fancyhead{}

\title{\name{}: ML-Generated Debugging Queries for Distributed Systems}

\pagestyle{plain}

\interfootnotelinepenalty 100000
\widowpenalty 100000
\clubpenalty 100000
\newfont{\tf}{phvro at 9.5pt}
\newfont{\tft}{phvro at 7.25pt}
\begin{abstract}

A major difficulty in debugging distributed systems lies in {\em manually} determining which of the many available debugging tools to use and how to query its logs. Our own study of a production debugging workflow confirms the magnitude of this burden. This paper explores whether a machine-learning model can assist developers in distributed systems debugging. We present \name{}, a debugging assistant which takes user reports and system logs as input, and outputs debugging queries that developers can use to find a bug's root cause. The key challenges lie in (1) combining inputs of different types (e.g., natural language reports and quantitative logs) and (2) generalizing to unseen faults.  \name{} addresses these by employing deep neural networks to uniformly embed diverse input sources and potential queries into a high-dimensional vector space. In addition, it exploits observations from production systems to factorize query generation into two computationally and statistically simpler learning tasks. To evaluate \name{}, we built a testbed with multiple distributed applications and debugging tools. By injecting faults and training on logs and reports from 800 Mechanical Turkers, we show that \name{} includes the most helpful query in its predicted list of top-3 relevant queries 96\% of the time. Our developer study confirms the utility of \name{}.
\end{abstract}

\settopmatter{printfolios=true}
\maketitle
\begin{sloppypar}

\section{Introduction}
\label{s:intro}

Developers often need to translate informal reports about problems provided by a user into actionable information that identifies the root cause of a bug. An ever-growing list of debugging tools aid developers in such root cause diagnosis. These tools enable a developer to log the behavior of applications running on end hosts~\cite{appdynamics, strace,opentracing,gdb,apachelogs,databaselogs}, end host networking stacks~\cite{tcpdump,bpf}, and network infrastructure~\cite{marple,snmp,int}. Some tools also track and relate execution across multiple subsystems of a distributed system~\cite{pivot,reverb,dapper,zipkin}.
Further, each tool allows developers to query the collected logs (e.g., using BPF expressions, SQL queries, or the graphical interfaces offered by interactive dashboards~\cite{grafana,datadog}) to hone in on interesting data. 

Yet, debugging distributed systems remains difficult, largely because it typically involves multiple {\em manual} steps~\cite{diff}: understand user reports,\footnote{We focus on user-generated reports, but note that the approach generalizes to auto-generated natural language crash reports, e.g., error messages.} iteratively issue debugging queries to test hypotheses about potential root causes, and finally develop a fix. As a concrete example, consider debugging a client interaction with a web service that results in an unusually slow page load. The developer receives a user report about the slow page load and has to develop a bug fix. However, the problem could be in many possible subsystems. At the client, the browser could be executing malformed HTML. At the server, there could be an unreachable database, a program error in the application frontend, or a misconfigured forwarding table in the network.

In the example above, the developer's difficulty is not the lack of tools; on the contrary, dozens of rich debugging frameworks exist for each subsystem (\S\ref{s:related}). Instead, the challenge lies in manually determining which debugging queries to issue, on which subsystems' logs, and with what parameters. Further, the answer to each question can depend on vast and heterogeneous logs. Indeed, distributed systems increasingly comprise many loosely coupled subsystems (e.g., microservices~\cite{lyft}), each with their own logging framework(s). As a result, developers face a significant cognitive burden to understand and correlate debugging information that exhibits heterogeneity in (1) data types (e.g., natural language user reports and error messages, numerical switch counters, RPC call graphs), (2) data sources (e.g., network infrastructure, end-host stack, application), and (3) abstraction levels (e.g., user reports, system-level logs, network-level counters).

Today, developers overcome these challenges using their hard-won intuition from debugging similar problems in the past, remembering which subsystems they investigated, and what debugging queries they issued. However, our developer survey and analysis of 4 months of debugging reports at a major SaaS company (\emph{Anon}) revealed that this manual approach consumes significant developer time (\S\ref{s:whyml}). At \emph{Anon}, moving from a user report to a root cause took developers an average of 8.5 hours, despite the fact that 94\% of the faults were repeated instances of the same type (e.g., resource underprovisioning), with only the fault location varying. Prior studies of other production systems have similarly observed the time significance of root cause analysis (relative to tasks such as triaging)~\cite{deepct, deeptriage}.

In this paper, motivated by the prevalence of large historical debugging datasets in software organizations~\cite{fbdata,ticketmasterdata} and the recurring nature of faults, we ask: {\em can a machine-learning (ML) model learn the same kind of developer intuition from past debugging experiences to accelerate root cause finding?} More precisely, given debugging data collected when a report was provoked, can an ML model automatically generate a debugging query that allows the developer to extract the most informative subset of the logs? Further, is the model's output ultimately useful: does it let the developer diagnose bugs faster than the status quo?

\begin{figure}[t]
\centering
\includegraphics[width=.45\textwidth]{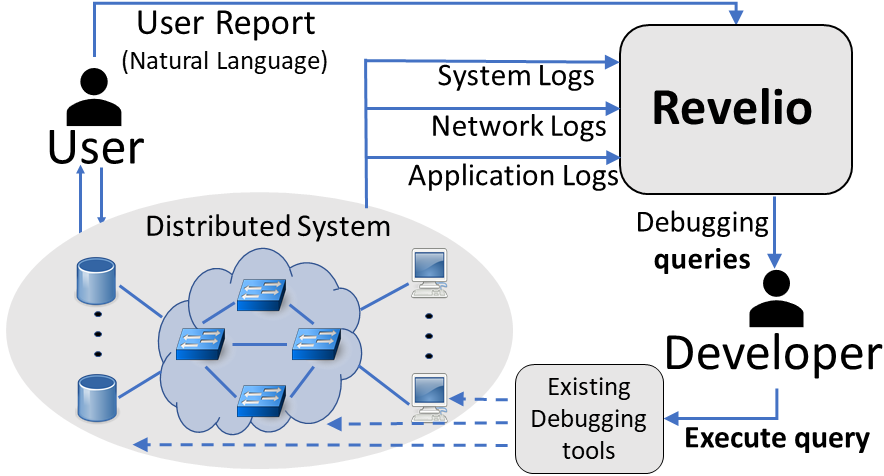}
\vspace{5pt}
\tightcaption{\name{} takes as input a user report and system logs, and outputs (for the developer) a ranked sequence of debugging queries that highlight the root cause. }
\vspace{5pt}
\label{fig:overview}
\end{figure}

To answer these questions, we developed \textbf{\name{}} (Figure~\ref{fig:overview}), whose goal is to help developers use existing debugging tools more effectively. \name{} takes as inputs user reports and system logs from existing tools, and outputs a ranked sequence of debugging queries that (when executed) elucidate the root cause. Note that \name{} is intended to be trained directly on the system logs being collected in a given deployment; we expect \name{} to be retrained if the set of tools or the data that they log changes.

\para{Why output debugging queries?} We chose to output a query sequence for several reasons. First, as evidenced by our production debugging analysis (\S\ref{ss:prod_work}), much developer time and effort in root cause analysis is spent selecting a subsystem to investigate, and determining how to use existing querying tools to analyze its collected logs -- queries embed both of these aspects. Second, bugs can often be tackled from multiple vantage points in distributed systems, e.g., congestion between two microservices can be resolved by either moving an application VM or changing inter-service routing rules. A sequence ensures that developers can extract root cause insights from those different vantage points to 

facilitate the generation of the appropriate fix. We next describe \name{}'s challenges and contributions.

\subsection{Challenges and Contributions}

\Para{1. Extensible model using distributed vector representations:} To combine \emph{heterogeneous data sources} (e.g., natural language user reports, numerical switch counters), we use neural networks that map each input to a high-dimensional distributed vector representation~\cite{code2vec, word2vec}, akin to intermediate representations like SSA~\cite{ssa} in programming languages. This makes our architecture {\em extensible}: a new type of debugging data can be incorporated by learning a mapping from that data type to a high-dimensional vector.

\Para{2. Modeling queries as vectors using graph convolutional networks:} Ideally, we should be able to convert queries into the same vector representation as our inputs; we could then find the relevance of a query to a particular debugging scenario by applying standard machine-learning concepts such as a similarity score between the query and input vectors. One approach is to simply assign a unique label to each query and employ a multi-label classifier to generate queries. However, this performs poorly (\S\ref{ss:extra}) because the opacity and independence of such labels fail to exploit the fact that debugging queries for a tool are all drawn from the same grammar. Instead, debugging queries are more faithfully modeled as abstract syntax trees (ASTs) in the syntax of the query language. To leverage this richer query format, we use \textit{graph convolutional networks} (GCNs)~\cite{kipf2016semi} to convert query ASTs into the same vector representation as our inputs.

\Para{3. Handling a large search space of queries using modularization:} For any inputs, the search space of potential queries is massive. This is because of the presence of many (1) \emph{query templates}, i.e., skeleton queries for a given subsystem with unspecified parameters, and (2) \emph{query parameters} that cover the scale of production systems (e.g., every IP address in the system could be a candidate value for a parameter). To handle this large search space, we {\em modularize} our ML model into two components. The first model uses user reports and system logs to \textit{predict a query template}; then the second model uses only the predicted template and system logs (not user reports), to \textit{predict numeric parameters}. This is motivated by our finding that production faults typically involve recurring types (\S\ref{s:whyml}), and can thus be debugged using a small set of templates---one per fault type. Modularization shrinks the output space of the first model, simplifying training computationally, and the input space of the second model, making it less likely to overfit to spurious input features.

\Para{4. Generalizing to unseen faults using abstraction:} To handle a large fraction of bugs in production settings, an ideal model should generalize to output useful queries for occurrences of previously seen bugs at new locations (\S\ref{s:whyml}). To achieve such generalization, we transform concrete switch/function ids into new, \emph{abstract} ids based on rank on some metric, e.g., queue size; consequently, the ids in one setup and another need not be the same, allowing us to generalize to new fault locations. For example, if the model captures a dependence on the largest router queue (which in the training set was Router $X$), it can generalize during testing to a different Router $Y$ with the largest router queue.

\subsection{Evaluating \name{}}

\Para{New distributed systems debugging testbed:} While organizations running distributed systems routinely collect the described debugging data, much of it is proprietary. To address this data scarcity and evaluate Revelio, we built a testbed (\S\ref{s:testbed}) on top of the Mininet emulation platform~\cite{mininet}. Our testbed currently integrates four debugging tools---Jaeger~\cite{jaeger}, Marple~\cite{marple}, cAdvisor~\cite{cadvisor}, and tcpdump~\cite{tcpdump}---and runs three industry-developed distributed applications---Reddit~\cite{reddit} (monolithic), Sock Shop~\cite{sockshop} (microservice), and Online Boutique~\cite{hipstershop} (microservice). In addition, our testbed includes an automatic fault injector that was informed by our analysis of production bugs (\S\ref{s:whyml}) and generates a variety of network, system, and application errors (\S\ref{ss:fault}). 

We enlisted 800 users on Amazon Mechanical Turk to interact with our testbed's applications. Turk users, unaware of the injected faults, were asked to report their experiences under these faulty scenarios via multiple choice and free-form questions. In total, we simulated 85 different faults (per app) and collected an average of 10 user reports per fault. We paired this training data with the system logs from our testbed, and a set of debugging queries generated by us to mimic those expected in debugging reports.

\Para{Testbed evaluation, developer study:} We evaluated \name{} using data from our testbed and Turk users (\S\ref{s:eval}), and with a developer study (\S\ref{s:devstudy}). Our key findings are: {\bf 1.} across the set of potential queries supported by our debugging tools, for repeat occurrences of the same faults, \name{} ranks the correct (i.e., the one that most directly highlights the root cause) query in the top-k 96\% (k=3), 100\% (k=4), and 100\% (k=5) of the time, {\bf 2.} \name{}'s model successfully generalizes to output the correct query 87\% (k=3), 88\% (k=4), and 100\% (k=5) of the time for faults that manifest in previously unseen locations, and {\bf 3.} developers with access to Revelio correctly identified 90\% of the root causes (compared to 60\% without \name{}), and did so 72\% (14 mins) faster. We additionally conducted quantitative experiments that demonstrate that each of our design choices, i.e., abstraction, GCNs, and modularization, individually has a significant (positive) effect on the query generation performance of \name{} (\S\ref{ss:extra}).

\section{\fontsize{10.5}{11}\selectfont{Production Bugs, Debugging Workflows}}
\label{s:whyml}

\begin{table*}[t!]
\footnotesize
\centering
\begin{tabular}{|p{3.4cm}|p{1.35cm}|p{1.75cm}|p{7cm}|p{2.12cm}|}
\hline
\textbf{Root Cause Category} & \textbf{\# of Tickets} & \textbf{\# of Locations} & \textbf{Example Root Cause} & \textbf{Avg. Diagnosis Time (mins)}\\
\hline
Resource underprovisioning & 17 & 11 & Load balancer is consuming all available memory and starving other co-located services & 293\\
\hline
Component failures & 58 & 29 & 3 nodes for a service were down, leading to queued 400 ERRORs & 176\\
\hline
Subsystem misconfigurations & 11 & 7 & Incorrect host mapping configuration in Zookeeper caused failure, and prevented cluster from servicing any events & 276\\
\hline
Network congestion & 5 & 4 & A spike in wide-area traffic caused unusually low data transfer rates between city1 and city2 & 725\\
\hline
Network-level misconfigurations  & 18 & 10 & Instances in a region are pointing to a NAT instance with incorrectly configured security groups, leading to dropped traffic & 92\\
\hline
Subsystem/Source-code bugs & 31 & 22 & Service returning 5xx errors due to a code change that added a condition on the availability of a parent asset ID & 1607\\
\hline
Incorrect data exchange & 26 & 16 & 4xx errors were being raised because the noise classifier service is sending additional data with each stock request & 417\\
\hline
One-off or unknown & 10 & 8 & 278 customer accounts were inadvertently canceled for unknown reason & 464\\
\hline
\end{tabular}
\vspace{15pt}
\tightcaption{Summary of closed debugging tickets at \textit{Anon} over a 4-month period. Examples have been partially anonymized and summarize the root causes listed in representative tickets.} 
\label{t:anon}
\end{table*}

To understand the operation and limitations of debugging tools and workflows in production distributed systems, we conducted a study at a major SaaS company (\textit{Anon}). Our analysis involved 7 services at \textit{Anon} that collectively handle 83 million user requests per day. Across these services, we examined the debugging process through a developer survey and a manual analysis of completed debugging tickets over a 4-month time period. To develop a general taxonomy for our analysis of \textit{Anon}'s data, we start with a literature survey of publicly reported bugs in production distributed systems.

\subsection{Literature Survey}
\label{ss:faults}

We surveyed many recent papers and blog posts that document or measure bugs in production settings. Our survey includes major outages in large-scale services (e.g., Dropbox~\cite{dropbox_outage}, Kubernetes~\cite{kubernetes_outage}), bugs in cloud services (e.g.,  Google~\cite{dapper}, Facebook~\cite{canopy}, Azure~\cite{azure_bugs}), and experiences with open source systems (e.g., Cassandra, HDFS~\cite{simpletesting}). Our survey revealed the following bug categories:

\begin{CompactEnumerate}
    \item \textbf{System software and configuration faults.}
    \squishlist
        \item \textbf{Resource underprovisioning~\cite{canopy,kubernetes_outage}}: In such bugs (e.g., at Facebook~\cite{canopy}), the containers or VMs running parts of a distributed system are allocated insufficient CPU, memory, disk, or network bandwidth. 

        \item \textbf{Component failures~\cite{xtrace, simpletesting, basecamp_outage, facebook_outage, dropbox_outage}}: Failures are common at scale, and can result from a faulty physical machine, a bug in the machine's hypervisor, or an unduly small amount of memory being allocated to a particular component. 
        
        \item \textbf{Subsystem misconfigurations~\cite{simpletesting, facebook_outage, azure_bugs}}: Errors in the internal configuration files for a given subsystem are common, especially given complex interoperation with other subsystems. Examples include incorrect hostname mappings that result in improper traffic routing and poorly configured values for timeouts or maximum connection limits~\cite{azure_bugs}.

    \squishend
    \item \textbf{Network faults.}
    \squishlist
        \item \textbf{Network congestion~\cite{canopy}}: Within data centers~\cite{canopy}, queues build up at various network locations (e.g., virtual and physical switches) that connect subsystems, either due to temporarily increased application traffic (e.g., TCP incast~\cite{incast}) or cross traffic. 
    
        \item \textbf{Incorrect network configuration~\cite{canopy,simpletesting,facebook_outage}}: Network devices (e.g., firewalls, NATs, switches) between subsystems that communicate via RPCs may be incorrectly configured with forward/drop rules.  This could cause unintended forwarding of packets to a destination or incorrect packet dropping.  
    \squishend
    \item \textbf{Application logic faults.}
    \squishlist
        \item \textbf{Bugs within subsystems~\cite{dapper, bugs_as_allergies, real_world_concurrency, parikshan, bugbench, instapaper_outage,kubernetes_outage}}: Bugs in application logic are prevalent in practice~\cite{reverb,diff}, and can result in a wide range of system effects. For example, certain bugs arise from (accidentally) inverted branch conditions that trigger seemingly inconsistent behavior: an application may traverse an incorrect branch and display incorrect content or result in a program error. In contrast, certain code changes can trigger performance degradations, e.g., if unnecessary RPC calls are generated between microservices. 
    
        \item \textbf{Incorrect data exchange formats and values~\cite{azure_bugs}}: Particularly in microservice settings as in Azure services~\cite{azure_bugs}, bugs can arise if the RPC formats of the sender and receiver do not match. For instance, a change in the API exposed by one microservice could result in a bug if its callers are unaware of this change. Also included in this category are certificate or credential updates that have only been partially distributed (resulting in access control errors).
    \squishend
\end{CompactEnumerate}

\subsection{Analysis of Debugging at \textit{Anon}}

\Para{Debugging workflow.}
\label{ss:prod_work}
Developers at \emph{Anon} use a variety of state-of-the-art monitoring tools (e.g., Splunk~\cite{splunk}, Datadog~\cite{datadog}, others~\cite{lightstep,newrelic,pingdom,icinga})  that continuously analyze system logs, visualize that data with dashboards, and raise alerts when anomalous or potentially buggy behavior is detected. These tools raise alerts based on either manually-specified heuristics and thresholds, or standard statistical analysis techniques that compare recent data to historical baselines (e.g., for outlier detection)~\cite{lightstep,prophet,donut,twitteranomaly}. As user- or internally-generated reports are filed, the burden of debugging falls largely to developers. For each report, developers must (1) filter through the raised alerts (across subsystems) to determine which are worth investigating and pertain to actual bugs and the issue at hand (vs. false positives), and (2) for bugs, find the root cause. Both steps involve iteratively analyzing low-level system logs, inspecting prior debugging tickets and the current report (both written in natural language), and issuing debugging queries (using query interfaces that run atop the same logs used to raise alerts~\cite{grafana,marple}). Once a root cause is identified, a summary of the issue, bug, and debugging process (e.g., investigated subsystems, issued queries) is documented as a completed debugging ticket.

\Para{Analysis of historical debugging tickets.} We manually analyzed all 176 debugging tickets that were created for the aforementioned services between November 2019 and February 2020. Our analysis involved manually clustering the tickets according to their root causes (as documented by \textit{Anon} developers). We selected clusters based on the fault categories extracted from our literature survey (\S\ref{ss:faults}); tickets that did not fall into one of these categories were placed in a ``One-off or unknown'' category. Table~\ref{t:anon} summarizes our findings, from which we make three primary observations:
\begin{CompactEnumerate}

\item There exists a small number of recurring categories of root causes that collectively represent the vast majority (94\%) of bugs.

\item The faults in a given category often manifest at different locations in the distributed system. For example, numerous ``Resource underprovisioning'' tickets involve high CPU loads but pertain to different servers, e.g., gateway servers vs. storage servers for popular data shards.

\item Identifying the root cause for a fault is time consuming, taking an average of 8.5 hours (min: 14 min, max: 2.9 days) across fault categories. We found that these lengthy durations are largely a result of the error-prone nature of root cause analysis: developers at \textit{Anon} must explore multiple subsystems (5 on avg.) and issue many debugging queries (8 on avg.) to find the root cause of a problem. Note that these debugging times are high even though most faults fall into recurring categories.

\end{CompactEnumerate}

\Para{Takeaways.} Our findings at \emph{Anon} collectively show that, while debugging tools have considerably improved (primarily for improved alert-raising and richer query interfaces), post-alert debugging, or moving from alerts to root causes of faults, is \emph{largely manual} and extremely \emph{time consuming}. These difficulties illustrate that existing monitoring tools and anomaly detectors commonly fail to elucidate the root cause of the problem, and still require much developer effort.  The focus of this paper is on automating the post-alert process for developers, i.e., ingesting diverse system logs \emph{and} natural language reports, and outputting debugging queries that highlight the root cause. To the best of our knowledge, there does not exist a solution for automating such query generation. Further, we believe that ML is the appropriate tool as it is unclear how to design automated heuristics that incorporate such diverse data sources and output full-fledged debugging queries (as opposed to, say, scalar alert thresholds). Additionally, the repetitive nature of faults also makes debugging amenable to a machine learning approach.

\Para{Goals and non-goals.} We note that our focus here is entirely on faults that fall into recurring categories---recall that such faults constitute the vast majority of faults at \textit{Anon}, and despite their recurring nature, still take significant time to diagnose. Importantly, we \emph{do not} target new fault categories and one-off faults that bear no similarity to prior ones, and instead leave debugging of those scenarios to future work.

\section{Overview of \name{}}
\label{s:design}

\begin{figure*}[!t]
\centering
\includegraphics[width=0.9\textwidth]{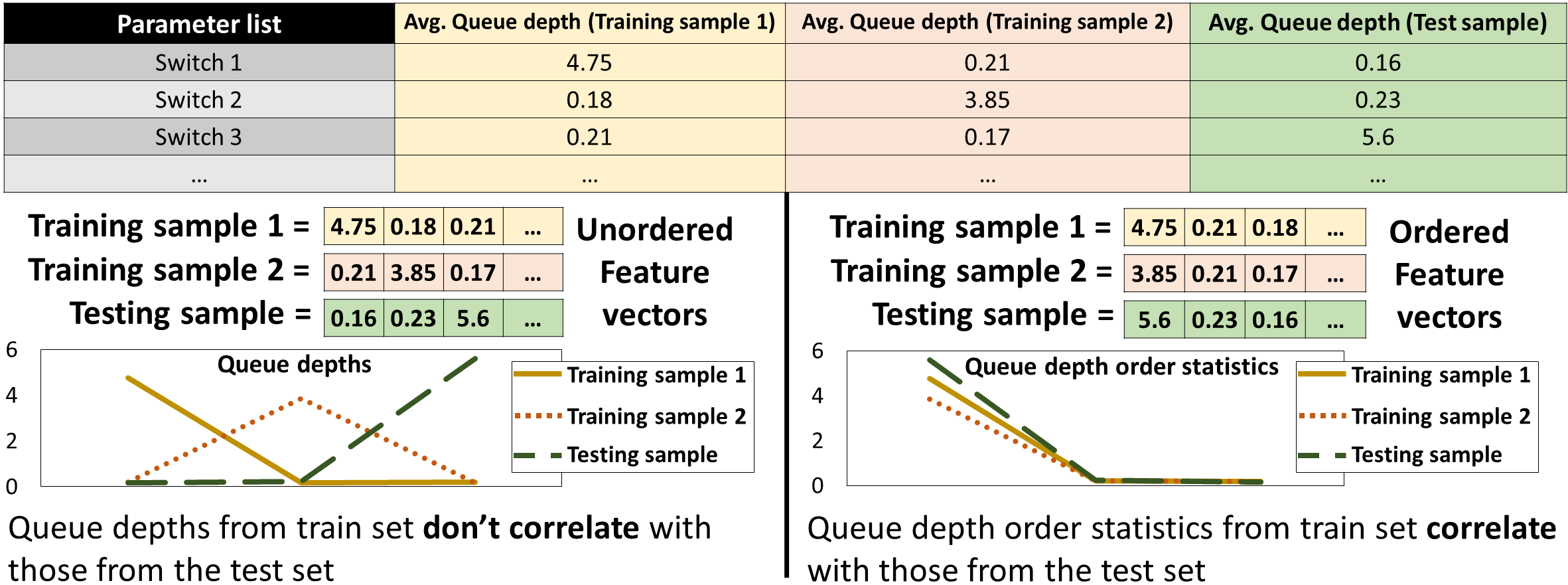}
\vspace{4pt}
\tightcaption{Example showing how rank-ordering helps to generalize to faults of the same type at different locations (e.g., switches 1 and 2 in training, switch 3 in testing). After ordering (right), despite the fault location being different, the queue depth order statistics in testing are correlated with those seen in training. In contrast, without ordering (left), the unseen fault location results in queue depth values that are quite dissimilar from training data.}
\vspace{8pt}

\label{fig:sort}
\end{figure*}

Informed by our findings at \textit{Anon}, we now present an overview of \name{}'s ML-based strategy to debugging query generation. We start with the challenges associated with an ML approach, and then describe the intuition behind our corresponding solutions; \S\ref{s:model} concretizes these insights by formally describing \name{}'s model.

At a high level, \name{} takes two inputs: (1) a \emph{user report} filed by a system user, and (2) the \emph{system logs} collected during the user's interactions with the system. The two sources provide distinct perspectives into the state of the system when a fault occurs---the former from an external and the latter from an internal viewpoint. Further, the two data sources differ fundamentally: system logs are highly structured, accurate, and contextually close to a developer's debugging options; user inputs are often noisy, unstructured (e.g., raw text), and abstract with respect to low-level system execution (e.g., a user may report that the system is slow to respond with no further information). As output, \name{} generates a ranked list of top-k debugging queries that are directly executable on the target debugging framework(s) (e.g., Jaeger~\cite{jaeger}) and highlight the root cause of the fault.

\subsection{Challenges}
\label{s:challenges}

\name{} must overcome four key challenges to generate debugging queries. First, the model has to combine and relate diverse and seemingly disparate data inputs. Second, the output space of queries is highly structured, making it harder than standard multi-label classification where each label is independent~\cite{bakir2007predicting}. This is because all debugging queries for a tool are drawn from the same language grammar, unlike opaque and independent labels. Third, the space of potential queries for a given input is large, requiring new techniques to scale to large distributed systems. Fourth, as per our study of production bugs (\S\ref{s:whyml}), the model must be able to generalize in a specific sense: if a fault occurs at one location during training and is debugged with a specific query, then, during testing, the model must predict the same query with a different parameter if the same fault occurs at a different location.

We now briefly describe how we handle each challenge before formally describing our model in \S\ref{s:model}

\subsection{Solutions}
\label{ss:solutions}

\begin{figure*}[t]
\centering
\includegraphics[width=0.85\textwidth]{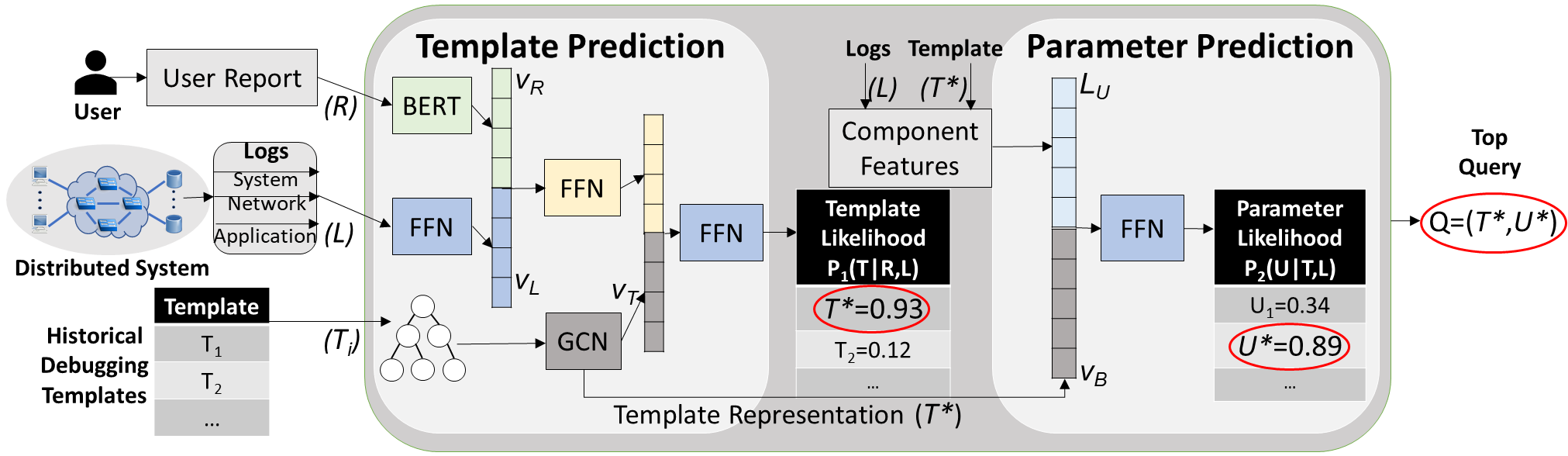}
\vspace{4pt}
\tightcaption{Overview of \name{}'s factorized, 2-phase approach to generating debugging queries for root cause diagnosis. For simplicity, the illustration assumes only a single debugging query as output (rather than a sequence of queries).}
\vspace{8pt}
\label{fig:model}
\end{figure*}

\Para{Challenge 1: Diverse data.} We handle diverse data sources by converting each into a vector and concatenating all vectors to form the system state vector. This has two benefits. First, each data source is normalized for downstream operations in the ML model. Second, the architecture is extensible: a new data source (e.g., crash reports) can be added by converting it into a vector (either learned or manually) that is then concatenated with the existing system state vector.

\Para{Challenge 2: Predicting queries.} To generate debugging queries, which can be represented as abstract syntax trees (ASTs) in the grammar of a tool's query language, we employ a Graph Convolutional Network to convert the AST into a query vector. A vector-based representation is easier to use with the rest of the ML model relative to richer representations such as trees. During training, given pairs of query and system vectors, we find model parameters that maximize the probability that these query vectors were predicted from these system vectors. During inference, given the ML model's parameters, we find the query that maximizes the probability of a query vector given the system vector.

\Para{Challenge 3: Scaling to large systems.} \name{} has to search over a large space of queries to output the best query in response to a given input. This search space scales with the size of the distributed system. To handle this, we exploit modularity and factorize our ML model into two cascaded components. The first uses user reports and system logs to generate query templates, which are skeleton queries for a particular subsystem with all numeric parameters left unspecified (e.g., SELECT \_ FROM \_). The second component then predicts the corresponding parameters using only the predicted template and system logs. This approach is motivated by two ideas. First, production faults typically involve recurring types (\S\ref{s:whyml}), and can thus be debugged using a small number of templates (one per fault type). Second, we assume that system logs sufficiently highlight the set of potential parameter values and the relative importance of each; as per \S\ref{s:challenges}, user reports are often abstract and rarely list parameter values (e.g., switch IDs). Modularization thus shrinks the output space of the first model, simplifying training computationally, regardless of system scale. It also shrinks the input space of the second model, making it less likely to overfit to spurious inputs, which in turn improves accuracy and generalizability.

\Para{Challenge 4: Generalizing to new fault locations.} Given the scale of production systems, it is infeasible to rely on training data that captures all possible locations of a given fault category.  Thus, our model should generalize to {\em different locations} for fault types seen during training. To aid with such generalization, we convert concrete switch/function ids in the system logs into abstract ids based on the rank order per feature (e.g., queue depth). This allows our models to learn the relevance of a given template or the importance of a particular subsystem based on a stable property like the subsystem's rank on a feature rather than a volatile property (e.g., switch ID); Figure~\ref{fig:sort} illustrates the utility of this approach. For example, during template prediction, the model is able to learn about the applicability of a template to the order statistics~\cite{order_statistics} of feature values across the system, rather than to the numerical or ordinal values of these features at specific subsystems. This is important because, if a given fault occurs at two different locations (both of which warrant the same template), the order statistics of feature values may be correlated, whereas the specific value assignments definitively will not. Similarly, for parameter prediction, ordering information is more robust to the addition, deletion, or restructuring of subsystems. 

\section{\name{}'s ML Model}
\label{s:model}

\begin{table}[!t]
\scriptsize
\centering
    \begin{tabular}{|p{0.75cm}|p{2.2cm}|p{4cm}|}
         \hline
         \textbf{Name} & \textbf{Description} & \textbf{Example} \\
         \hline
         $T$ & Query template & SELECT QUEUE\_SIZE FROM T WHERE SWITCH\_ID = \_\\
         \hline
         &Blanks in template&\\
         $B$ & $\{b_1,b_2,...,b_z\}$ & $b_i$ = \_ in the above example \\
         \hline
         &Query parameters&\\
         $U$ & $\{u_1,u_2,...,u_z\}$ & $u_i$ = switch ID\\
         \hline
         $R$ & User report & ``Page is loading slowly'' \\
         \hline
         $L$ & System logs & OpenTracing and Marple logs \\
         \hline
    \end{tabular}
    \vspace{15pt}
    \tightcaption{Variables in Revelio's ML model. Figure~\ref{fig:revelio_inputs} in \S\ref{s:appendix} lists example input values for each.}
    \label{tab:revelio_ml}
\end{table}

To enable Revelio's prediction capabilities, we need to induce a distribution (from data) $$\Prob(Q | R, L)$$ where $Q$ is a debugging query, $R$ is a user report, and $L$ refers to the system logs (Table~\ref{tab:revelio_ml} lists the variables in our model). Once the parameters of this distribution have been learned by maximum likelihood, the distribution allows us to predict the query $Q$ that maximizes $\Prob(Q | R, L)$. The data we require for this is a set of triples $\langle R, L, Q\rangle$. While the above formulation seems straightforward at first glance, it involves learning a probability distribution over all possible queries and across all tools, which is extremely challenging and requires a substantial amount of data. Therefore, we instead split up each query $Q$ into a query template $T$ (e.g., SELECT \_ FROM \_) and a set of values $U$ (to fill in the blanks). This allows us to factorize the previous distribution as:
\begin{small}
\begin{equation}
\Prob(Q | R, L) = \Prob(T, U | R, L) = \Prob_1(T | R, L) \Prob_2(U | T, R, L)    
\end{equation}
\end{small}
To simplify our training further, we make an independence assumption on $P_2$  by assuming that $R$ is not likely to help predict $U$ (as described in \S\ref{ss:solutions}). Thus, we have:

\begin{small}
\begin{equation}
\Prob_2(U | T, R, L) = \Prob_2(U | T, L)    
\end{equation}
\end{small}

We can further factorize this into a product of distributions over values $u_i$ for each blank $b_i$ in the template $T$:
\begin{small}
\begin{equation}
\Prob_2(U = \{u_1, u_2, ..., u_z\}| T, L) = \prod_{i \in [1, z]}{\Prob_2( u_i | b_i, T, L)}
\end{equation}
\end{small}

where $z$ is the total number of blanks in the template.

From an inference standpoint, this means we have a 2-phase query generation process: we first generate a query template and then fill in the blanks with appropriate values using the system logs (Figure~\ref{fig:model}). We next detail how we model each of the distributions ($\Prob_1$ and $\Prob_2$), as well as our learning and inference procedures for each.

\subsection{Predicting Probabilities for Query Templates ($\Prob_1$)}
\label{s:model1}

Assume the user report $R$ to be in the form of raw text and $L$ to be a vector obtained by concatenating ordered vectors for each feature (Figures~\ref{fig:feature_sorting} in \S\ref{s:appendix}) extracted from the system logs (e.g., time-windowed average, min queueing delay). Recall from \S\ref{ss:solutions} that rank ordering per feature in $L$ enables our model to learn about the order statistics of feature values across subsystems, rather than about numerical or ordinal values at specific subsystems (Figure~\ref{fig:sort}). From here, a straightforward way of modeling $\Prob_1(T | R, L)$ would be to use a multi-label classifier with each template $T$ being a different label. However, as discussed in \S\ref{s:design}, query templates are structured and made up of smaller atomic components (e.g., IF, MAX, MEAN statements). In other words, the ASTs of many query templates share common subtrees. Therefore, \textit{simply treating each template as an independent output label is wasteful in terms of not sharing statistical strength}.

Therefore, we adopt a different approach to modeling the output templates. In order to preserve the structural aspects in queries, we represent each template $T$ in the form of an abstract syntax tree (AST). Each node in the tree is an operator (e.g., SELECT) and the edges represent how the operators are composed together to form larger trees.

We use a Graph Convolutional Network (GCN)~\cite{kipf2016semi} to construct a vector representation $v_T$ for each query template's abstract syntax tree. The GCN updates each node's vector representation in the AST by pooling information from all its neighbors and performs this process multiple times, allowing it to combine information from all nodes in the tree. The GCN outputs a vector for each node in the tree -- we take the vector of the root node  $v_T$ to represent the tree's information. In parallel, we use a contextual text encoder (BERT)~\cite{bert} to convert the issue report $R$ into a vector $v_R$ and pass the log $L$ through a linear neural network layer to get a vector $v_L$. $v_R$ and $v_L$ are concatenated and fed through a non-linear layer followed by a linear layer to get a single vector $v_S$ representing the system state from both internal and external viewpoints. Finally, we use both $v_S$ and $v_T$ to obtain a measure for how likely the template $T$ is applicable to the debugging scenario $\langle R, L \rangle$ (i.e., the probability of $T$ given $R$ and $L$). We then search for a set of neural network parameters that maximize this score (S) or likelihood.
The sequence of operations are summarized as:

\begin{small}
\begin{eqnarray*}
v_T &=& \textsc{GCN}(T)[root] \\
v_R &=& \textsc{BERT}(R) \\
v_L &=& \textsc{Linear}(L) \\
v_S &=& \textsc{Linear}(\textsc{ReLU}([v_R; v_L; v_T])) \\
S(T, R, L) &=&  \textsc{Linear}(\textsc{ReLU}([v_S; v_T]))\\
\Prob_1(T | R, L) &=& \textsc{Softmax}(S(T, R, L)) = \frac{S(T, R, L)}{\sum_{T'} S(T', R, L)}
\end{eqnarray*}
\end{small}

where $[;]$ represents a concatenation of two or more vectors and $GCN(T)[root]$ represents indexing the output of the GCN to get the vector of the root node.

All of the above operations represent a continuous flow of information through a \emph{single} deep neural network whose parameters $\theta$ can be trained through back-propagation and stochastic gradient descent~\cite{goodfellow2016deep}.
We use the following maximization objective to learn the parameters:
\vspace{-0.8pt}
\begin{small}
\begin{equation}
\max_\theta \mathcal{L}(\theta) = \sum_{(T, R, L) \sim \mathcal{D}} \Prob_1(T | R, L)
\label{eq:obj1}
\end{equation}
\end{small}

Enumerating all trees $T'$ is intractable, so we employ Noise Contrastive Estimation (NCE)~\cite{gutmann2010noise} and draw $m=2$ negative samples to form each $T'$ to approximate the objective.

\subsection{Predicting Values to Fill Query Templates ($\Prob_2$)}
Now that we have a method to pick a template $T^*$, we need to fill in the values for each blank $b$ in $T^*$.\footnote{For ease of exposition, we will assume filling in a single value, but our method can easily be used to fill in multiple values, one at a time.}
Each template implicitly specifies the type of subsystem (e.g., switches for a Marple query) that is relevant for the fault at hand. Thus, using the template, we first extract a list of all relevant subsystems from the system logs $L$. For each subsystem $u$ in this list, we have a feature vector $L_u$ which summarizes all of its logs (e.g., avg/max queue depth, packet count; see Table~\ref{t:featurelist} for a full list). We also include ranking information $rank_u$ for each feature in $L_u$ (e.g., $u$'s rank in queue depth across all switches). Note that ranks embed the same information as ordering from \S\ref{s:model1}; ordering is not possible here because each $L_u$ pertains to only a single subsystem. We use these features, along with a vector representation of the blank in the template (described below), to pick the most likely subsystem for the blank. 

We feed the template $T^*$ (represented as an AST) through the same GCN module as in \S\ref{s:model1} and choose the vector representation for blank $b$ to be the output vector of its corresponding node in the tree. This allows us to represent the requirements of $b$ using the properties of its neighboring nodes in the AST. Our goal is to then pick the \emph{most suitable subsystem $u$} for the blank, and return the corresponding system identifier (e.g., IP address or port number). We use a similar set of operations to those in \S\ref{s:model1} to pick the most likely $u$ to fill $b$:

{\fontsize{9}{12}\selectfont
\begin{eqnarray*}
v_b &=& \textsc{GCN}(T)[b] \\
S(u, b, T, L) &=& \textsc{Linear}(\textsc{ReLU}(\textsc{Linear}( v_b; L_u; rank_u)))\\
\Prob_2(u | b, T, L) &= & \textsc{Softmax}(S(u, b, T, L))
\end{eqnarray*}}

where $rank_u$ indicates the rank of subsystem $u$ in its subsystem's logs $L$, based on the feature of interest (e.g., rank of a switch, across all switches, based on mean queue depth).
We then use an objective similar to Eq.~\ref{eq:obj1} to maximize $\Prob_2$ over ground truth data and learn the model parameters $\phi$.

\subsection{Choosing the Final Queries}
Once each of the two models above have been trained, during inference, we find the combination of query template and query parameters that maximizes the probability that the resulting query would result from the given system state vector. This probability in turn is the product of the two probabilities predicted by each of our models $\Prob_1$ and $\Prob_2$ above. 
\begin{equation}
    Q^* = (T^*, U^*) = \arg\max_{T, U} \Prob_1(T | R, L) \Prob_2(U | T, L)
\end{equation}
We can also pick the top $k$ most relevant queries, rather than just the single most relevant one, using the ranking produced by the probabilities above. If $|T|\times|U|$ proves to be very large, we can approximate the above computation by considering only the top few templates according to $\Prob_1 (T | R, L)$.

\subsection{Implementation details}

For all FFN layers in our ranking model, we use two linear layers, each with hidden size 300, along with ReLU non-linearity. The GCN also uses a hidden vector size of 300. We use the Adam optimizer~\cite{kingma2014adam} with a learning rate of 0.0001. 

\section{Systems Debugging Testbed}
\label{s:testbed}

Developing and testing \name{} requires access to an operational distributed systems environment with source code, application data, debugging data, and integrated debugging tools. Industrial systems satisfy these requirements and have large amounts of debugging data internally~\cite{fbdata, ticketmasterdata}. Unfortunately, to the best of our knowledge, no such environment exists for public use. While we were able to analyze debugging reports at \textit{Anon} (\S\ref{s:whyml}), we could not access raw system logs, precluding the use of \name{} at \textit{Anon}.

\begin{figure}[t]
\centering
\includegraphics[width=0.4\textwidth]{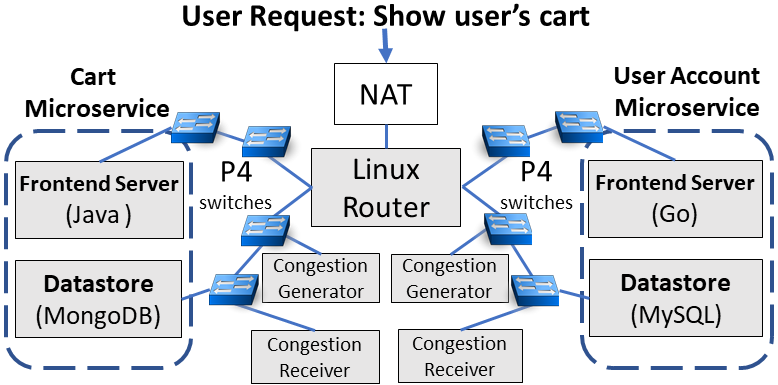}
\vspace{5pt}
\tightcaption{A slice (2/14 microservices) of the topology of our distributed systems testbed for Sock Shop~\cite{sockshop}; Reddit~\cite{reddit} and Online Boutique~\cite{hipstershop} followed the same pattern and are illustrated in \S\ref{s:appendix}. Each switch has a congestion generator/receiver, and the testbed incorporates four debugging tools and a fault injection service (all omitted for space).}

\label{fig:testbed}
\end{figure}

In this section, we describe the testbed (Figure~\ref{fig:testbed}) that we developed to fill this void. Our current implementation integrates multiple features to mimic realistic debugging cases. In particular, it incorporates three open-source distributed applications~\cite{reddit, sockshop, hipstershop},
four state-of-the-art systems and network debugging tools~\cite{opentracing, marple}, and an automatic fault injection service to generate production-like debugging scenarios. Remote users can interact with the applications in our testbed using a web browser and report any performance or correctness issues they observe. The setup is structured to be compatible with most distributed applications, and can be easily extended to incorporate new tools~\cite{canopy,xtrace,appdynamics_query}.

\subsection{\fontsize{9.5}{10}\selectfont Single-Machine Emulation of Distributed Apps}

\para{Overview of applications.} Our current testbed implementation considers three different distributed web applications, which involve both monolithic and microservice-based architectures: Reddit~\cite{reddit} (monolithic), Sock Shop~\cite{sockshop} (microservice-based), and Online Boutique~\cite{hipstershop} (microservice-based). For each application, we use the publicly available source code that was provided by the corresponding industrial organization and is intended to capture the technologies and architectures that they employ in their production services. 
\S\ref{ss:apps} provides additional details. 

Our goal is to run each application in a distributed and controlled manner, in order to scale to large workloads and deployments, consider broad sets of realistic distributed debugging faults, and ultimately generate complete debugging datasets for \name{}.

One approach would be to run each application service on VM instances in the public cloud. However, public cloud offerings typically hide inter-instance network components (e.g., switches, firewalls) from users, precluding the use of in-network debugging tools (\S\ref{s:related}).

Instead, we opt for a local emulation approach in which we run each application subsystem (or service) in a different container on the same physical machine, and specify the network infrastructure and connectivity between them. To do this, we use Containernet~\cite{containernet}, an extension of Mininet~\cite{mininet} that can coordinate Docker~\cite{docker} containers, each running on a dedicated core; we assign a separate core for network operation (i.e., P4 switch simulation). Our testbed can be scaled up to support higher throughput by making use of additional physical machines through distributed emulation~\cite{maxinet}.

As illustrated in Figure~\ref{fig:testbed}, for each application, we configure its subsystem/service containers into a star topology.

At the center of the topology is a router which is responsible for layer 3 faults (e.g., firewall configuration errors). Each subsystem is connected to this router via two P4 programmable switches~\cite{p4_behavioral} using Mininet. We set routing rules to ensure that all subsystems are appropriately reachable. Finally, for external reachability, a NAT is used to connect the central router to the host machine's Internet-reachable interface.

\subsection{Integrating Debugging Tools}

We integrate four debugging tools into our setup:
\squishlist

\item \textbf{Marple}~\cite{marple} is a performance query language for network monitoring that uses SQL-like constructs (e.g., groupby, filter) to support queries that track 1) per-packet and per-switch queuing delays, and 2) user-defined aggregation functions across packets.

\item \textbf{tcpdump}~\cite{tcpdump} is an end-host network stack inspector which analyzes all packets flowing through the host's network interfaces, and supports querying in the form of packet content filtering (e.g., by hostname, checksum).

\item \textbf{Uber's Jaeger}~\cite{jaeger} is an end-to-end distributed systems tracing system which follows the OpenTracing specification~\cite{opentracing}. With Jaeger, developers embed tracepoints directly into their system source code to log custom state (e.g., variable values), and then aggregate tracepoint and timing information to understand how data values and control state flow across time and subsystems.

\item \textbf{Google's cAdvisor}~\cite{cadvisor} profiles the resource utilization of individual containers, logging the following every 1 second: instantaneous CPU usage, memory usage, disk read/write throughput, and cumulative number of page faults.

\squishend
\S\ref{ss:tools} presents more details regarding the integration and usage of each tool. We note that these tools represent only part of the state-of-the-art for network and distributed systems monitoring; our setup is amenable to others~\cite{canopy,xtrace,appdynamics_query}.

\subsection{Fault Injection Service}
\label{ss:fault}

To create debugging data from realistic debugging scenarios, we created an automatic fault injection service. We note that our goal is not necessarily to match the system scale at which production faults were reported, but instead to evoke the user reports, system log patterns, and queries that correspond to the reported fault categories.

Our service is guided by our literature survey of production faults and our findings at \textit{Anon} (\S\ref{s:whyml}). Specifically, we incorporate faults that cover all of the categories discussed in \S\ref{s:whyml}, and match the ratios across categories with the data from \textit{Anon} (Table~\ref{t:anon}). These categories cover the observable performance (i.e., increased system response times) and functionality (i.e., missing or inconsistent page content, crashes) issues for the applications we consider. Table~\ref{t:fault_list} (in \S\ref{s:appendix}) provides more detailed examples of the specific faults we inject. We note that our service can be easily extended to incorporate new bug types.

\para{Injecting faults:} Our fault injector operates differently per fault class. For network or system configuration faults, we use Mininet and Docker commands to bring down a subsystem, start a congestion generator, change a service's provisioned resource values, or inject a firewall/routing rule at the router. In contrast, application logic faults require modified source code. In each container pertaining to an application's service logic, we include a script that takes in a fault instruction and replaces the appropriate source code with a version embedding the fault, and restarts the application.

\section{Evaluation}
\label{s:eval}

\subsection{Data Collection}
\label{ss:methodology}

\begin{table}[t!]
\footnotesize
\centering
\begin{tabular}{| c | c | c | c |}
\hline
\textbf{Metric} & \textbf{Reddit} & \textbf{Sock Shop} & \textbf{Online Boutique}\\
\hline
\# of Unique Faults & 76 & 102 & 80\\
\hline
\# of Unique Queries & 118 & 320 & 269\\
\hline
Query Vocabulary Size & 60 & 136 & 122\\
\hline
Report Vocabulary Size & 1040 & 1327 & 1258\\
\hline
\end{tabular}
\vspace{15pt}
\tightcaption{Summary of debugging queries and Turk user reports.}
\vspace{3pt}
\label{t:dataset_stats}
\end{table}

To extract system logs, user reports, and debugging queries from our testbed (\S\ref{s:testbed}), we conducted a large-scale data collection experiment on Amazon Mechanical Turk. For each application, we set up an EC2 instance per fault that we consider (Table~\ref{t:dataset_stats}). Each instance runs the entire testbed for that application, with the associated fault injected into it. All instances for an application were populated with the same content, which was generated using an application-provided script, e.g., Reddit content was scraped from the live site. %

Our experiment supported only ``master'' Turk users, and each was only allowed to participate once per fault+application pair. Each user was assigned to a specific instance/fault at random, and was presented with a UI (Figure~\ref{fig:ui} in \S\ref{s:appendix}) that had an iframe pointing to the corresponding instance's frontend web server. Users were asked to perform multiple tasks within each application, including loading the homepage, clicking on item pages or user profiles, adding comments, and adding items to their carts. Prior to the experiment, users were shown representative pages for each step (to ensure familiarity with the interfaces), and told about expected bug-free load times (4-5 sec in our setup).


\begin{table}[t!]
\footnotesize
\centering
\begin{tabular}{ c | c| c}
\hline
\textbf{Marple} & \textbf{Jaeger} & \textbf{cAdvisor} \\
\hline
Packet count & \# of accessed variables & CPU utilization \\
\hline
Queue depth & Duration of execution & Memory utilization \\
\hline
N/A & \# of exceptions thrown & Disk throughput \\
\hline
\end{tabular}
\vspace{15pt}
\tightcaption{Metrics in system logs. Marple, Jaeger, and cAdvisor metrics are recorded per-switch, per-function, and per-container; tcpdump is omitted for space. We featurize each metric using standard summary statistics (e.g., avg, stdev, max).}
\vspace{-3pt}
\label{t:featurelist}
\end{table}

For each task, users were asked to report performance and functionality issues into a form that included both multiple choice and free-form questions. During each user's experiment, the standard system logs for each tool in our testbed (\S\ref{ss:tools}) were collected on the instance; Table~\ref{t:featurelist} lists the collected metrics. We condense and featurize the time-series data for each metric using standard summary statistics such as min, max, average, and median. Once a user completed the experiment, their reports were paired with the associated system logs. We allowed up to 5 concurrent users per instance, and system logs reflect the interactions of all concurrent users. To complete our dataset, for each fault, we generate a debugging query with the appropriate tool that sufficiently highlights the root cause for the fault. This query is intended to represent the result of a past (successful) debugging experience. Table~\ref{t:dataset_stats} summarizes our dataset, and Table~\ref{t:report_examples} (in \S\ref{s:appendix}) lists example user reports.

\para{Methodology:} We divided the dataset for each application into 53\% for training, 13\% for validation, and 34\% for testing. We further divided our testing data into two test sets, \emph{test\_generalize} and \emph{test\_repeat}. \emph{test\_generalize} evaluates \name{}'s ability to generalize to new locations for previously seen fault types, and includes only data for faults that have matching query templates in the training data, but different parameters. \emph{test\_repeat} evaluates \name{}'s ability to suggest relevant queries for repeat faults, and includes only data for faults that have matching query templates \emph{and} parameters in the training data. All presented results test the best observed model from the validation set on the test sets. We evaluate \name{} primarily using two metrics: 1) \textit{rank} of the correct query (i.e., the query which most directly highlights the root cause) among the ordered list of the model's predicted queries, and 2) \textit{top-k accuracy}, which we define as the presence of the correct query in the top-k predictions.

\para{Result presentation:} Results for Online Boutique are similar to the other two applications, but omitted for space; deep-dive results (\S\ref{ss:extra}) are for Reddit, but trends hold for all three.

\newcommand{\PreserveBackslash}[1]{\let\temp=\\#1\let\\=\temp}
\newcolumntype{C}[1]{>{\PreserveBackslash\centering}p{#1}}
\begin{table*}[t!]
\footnotesize
\centering
\begin{tabular}{|C{8.5cm}|C{8.5cm}|}
\hline
 \textbf{Ground Truth Query} & \textbf{\name{}'s Top-Ranked Predicted Query} \\
 \hline
\texttt{stream = filter(T, switch==3)} & \texttt{SELECT span FROM spans WHERE name="GET\_comments"} \\
\texttt{result = groupby(stream, [5-tuple], count);}& \textit{(This \textbf{Jaeger} query also helps identify the Memcache failure by}\\
\textit{(This \textbf{Marple} query highlights the lack of network traffic to/from a (failed) Memcache instance.)} & \textit{honing in on the tracepoint with the corresponding 'connection failed' error message.)} \\
\hline
\texttt{SELECT * FROM cpu\_usage WHERE host="mn.h1"} & \texttt{SELECT span FROM spans WHERE name="\_byID"} \\
\textit{(This \textbf{cAdvisor} query helps to identify that a given host} & \textit{(This \textbf{Jaeger} query also helps highlight the host's underprovisioned} \\
\textit{is consistently running at 100\% CPU utilization, and is thus underprovisioned).} & \textit{CPU resources by showing the ensuing high function execution times on the host.)} \\
\hline
\texttt{SELECT span FROM spans WHERE name="\_find\_rels"} & \texttt{stream = filter(T, switch==1)} \\
\textit{(This \textbf{Jaeger} query helps to identify that a bug in} & \texttt{result = groupby(stream, [5-tuple], count);} \\
\textit{a function is resulting in no queries being issued to a MongoDB database.)} & \textit{(This \textbf{Marple} query shows the lack of network traffic between the host of a buggy function and the database.)} \\
\hline
\end{tabular}
\vspace{14pt}
\tightcaption{Examples where \name{}'s top-ranked predictions do not match the ground truth queries.
In all cases, \name{}'s top query is highly relevant, but characterizes the fault from an alternate system vantage point. Queries are condensed for ease of disposition.}
\label{t:error_analysis}
\end{table*}

\begin{figure}[t]
\centering
\includegraphics[width=.37\textwidth]{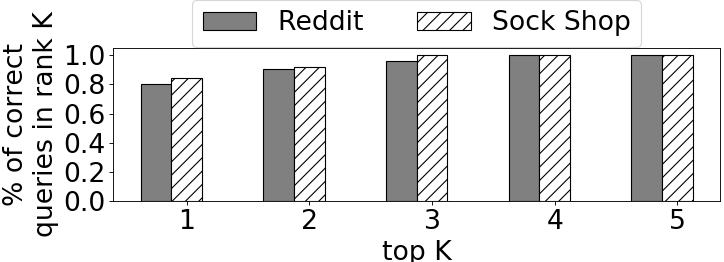}
\vspace{3pt}
\tightcaption{Cumulative distribution (per app) of the rank of the correct query over our test set of \emph{repeat} faults.}
\vspace{2pt}
\label{fig:repeat}
\end{figure}

\subsection{Evaluating \name{}'s Queries}

\para{Repeat faults.} For each fault in test\_repeat, we measured the rank of the correct query in \name{}'s predictions. As shown in Figure~\ref{fig:repeat}, for 80\% of the test samples for Reddit, \name{} assigns a rank of 1 to the correct query. Further, for 96\% of the Reddit test cases, the correct query is within the top 3 predicted queries. Performance is similar for Sock Shop, with the correct query being in the top 3 100\% of the time.

\begin{figure}[t]
\centering
\includegraphics[width=.37\textwidth]{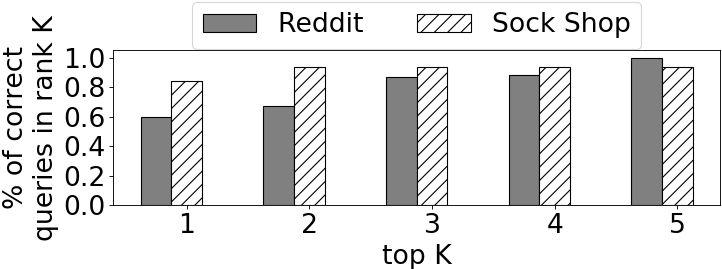}
\tightcaption{Cumulative distribution (per app) of the rank of the correct query over our test set of \emph{previously unseen} faults.}
\label{fig:generalize}
\end{figure}

\para{Generalizing to new fault locations.} Figure~\ref{fig:generalize} shows results for the more challenging scenario of new fault locations for repeat fault types (i.e., test\_generalize). As shown, \name{} is still able to consistently predict the correct query: \name{}'s model assigns a rank of 1 to the correct query 60\% and 85\% of the time for Reddit and Sock Shop. For both applications, the correct query was \emph{always} in the top-5 predictions.

\para{Benefits of query sequences:} To gain further insights into \name{}'s predicted queries, we analyzed scenarios in which the correct query was not ranked as 1. We find that in these cases, despite not matching the ground truth, \name{}'s highly ranked queries typically relate to the fault at hand, but characterize it from different vantage points.

Table~\ref{t:error_analysis} lists three representative examples. For instance, the first example pertains to a fault in which the Memcache subsystem is down for Reddit. The correct query is a Marple query which tracks packet counts at the switch directly connected to the failed subsystem; this query would highlight the lack of incoming/outgoing network traffic from Memcache. However, \name{}'s top-ranked query was a Jaeger query which hones in on a tracepoint that contains an error message noting the inability to connect to Memcache.

Similarly, in the second example, a Sock Shop subsystem is not provisioned sufficient CPU resources to handle the incoming traffic. The correct query was a cAdvisor query that explicitly tracked the container's CPU usage, but \name{}'s top-ranked query used Jaeger to track the high residual function execution times for the microservice running in that container.

In cases where \name{}'s top-ranked query is not the correct query, \name{} most often ranked the correct query as second or third. Thus, by outputting a sequence of ranked debugging queries, \name{} can provide developers with significant context about a fault from multiple vantage points.

\subsection{Understanding \name{}}
\label{ss:extra}

\begin{table}[t!]
\centering
\small
\begin{tabular}{|c|c|c|}
\hline
\textbf{Scenario} & \textbf{test\_repeat} & \textbf{test\_generalize} \\
\hline
User report+system logs & 1.33 (100\%) & 1.97 (100\%)\\
\hline

Only system logs & 1.86 (100\%) & 2.29 (90.2\%)\\
\hline
\end{tabular}
\vspace{14pt}
\tightcaption{Impact of different input sources on \name{}'s performance. Results list avg rank (\% in top-5) and are for Reddit.}
\vspace{1pt}
\label{t:sources}
\end{table}

\para{Importance of user reports:}
By default, \name{}'s model accepts both natural language user reports and quantitative system logs. To understand the importance of considering user reports in query generation, we evaluated a version of \name{} that excludes user reports from its input set; note that system logs cannot be excluded as they are required for parameter prediction. As shown in Table~\ref{t:sources}, \name{} significantly benefits from having access to both inputs. For example, on test\_generalize, the average rank of the correct query is 1.97 and 2.29 with and without user reports, respectively.

\begin{figure}[t]
\centering
\includegraphics[width=.35\textwidth]{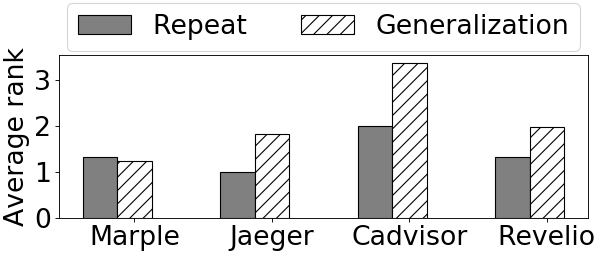}
\vspace{9pt}
\tightcaption{Comparing the average rank for single-tool and multi-tool versions of \name{}. Results are for Reddit.}
\vspace{6pt}
\label{fig:ablation}
\end{figure}

\para{Multi-tool vs. single-tool models:} We performed another ablation study where we compare \name{} when training and testing on logs from Marple, Jaeger, cAdvisor, and tcpdump together (multi-tool model), and in isolation (single-tool model). For each isolated tool, we prune the training, validation, test\_repeat, and test\_generalize sets to include only faults pertaining to that tool. As shown in Figure~\ref{fig:ablation}, we find that the per-tool models achieve better average ranks than the combined (default) model. The reason is that focusing on one tool allows \name{} to predict templates and parameters from a smaller space. However, our results show that \name{} pays only a small cost for operating across debugging tools: the average rank in the combined model is only 33\% higher than the best per-tool model for test\_repeat. This is key to \name{}'s ability to alleviate the burden of determining which tool to use for a particular scenario.

\para{Model structure:} To understand the importance of \name{}'s model structure and composition (\S\ref{s:model}), we compared it with the following variants that each modified one key design choice: 1) \textit{Revelio\_monolithic} uses a single model (not factorized) to output a fully formed query, 2) \textit{Revelio\_no\_rank\_order} eliminates the rank ordering of features in \name{}'s template and parameter prediction models, and 3) \textit{\name{}\_classifier} uses a multi-label classifier to select query templates rather than employing a GCN to construct a vector representation of each template's AST. Table~\ref{t:models} lists our results which highlight three main points. First, \name{} outperforms \textit{Revelio\_monolithic} on both test sets, highlighting the importance of factorization in terms of simplifying (both computationally and statistically) the difficult task of query prediction, particularly for generalization. Second, by rank ordering feature values, \name{} achieves an average rank of 1.97 for test\_generalize; in contrast, \textit{Revelio\_no\_rank\_order} is fundamentally unable to predict templates and parameters (and thus, queries) for repeat fault types in new (i.e., unseen during training) locations. Third, \name{}'s improved performance over \textit{\name{}\_classifier} illustrates the importance of using a GCN to extract semantic information about query structure (which a classifier cannot).

\begin{table}[t!]
\centering
\small
\begin{tabular}{|c|c|c|}
\hline
\textbf{Model} & \textbf{test\_repeat} & \textbf{test\_generalize} \\
\hline
\name{} & 1.33 (100\%) & 1.97 (100\%)\\
\hline
\name{}\_monolithic & 17.5 (15.1\%) & 22.4 (18.5\%)\\
\hline
\name{}\_no\_rank\_order & 1.29 (100\%) & N/A \\
\hline
\name{}\_classifier & 2.41 (88.7\%) & 2.69 (86.9\%)\\
\hline

\end{tabular}
\vspace{15pt}
\tightcaption{Impact of \name{}'s modularization rather than one model (\_monolithic), use of rank-ordering (\_no\_rank\_order), and use of a GCN rather than a multi-label classifier (\_classifier). Results list avg rank (\% in top-5) for Reddit.}
\vspace{0pt}
\label{t:models}
\end{table}

\begin{table}[t!]

\centering
\small

\begin{tabular}{|c|c|c|}
\hline
\textbf{Removed Feature} & \textbf{test\_repeat} & \textbf{test\_generalize} \\
\hline
\cellcolor{blue!12}Packet count & 2.14 (100\%) & 7.93 (80.4\%)\\
\hline
\cellcolor{blue!12}Queueing delay & 2.27 (96.1\%) & 2.51 (92.4\%)\\
\hline
\cellcolor{red!32}Variable count & 1.67 (96.1\%) & 3.54 (71.7\%) \\
\hline
\cellcolor{red!32}Duration of execution & 7.29 (88.2\%) & 4.11 (89.1\%) \\
\hline
\cellcolor{gray!23}CPU utilization & 1.65 (96.1\%) & 4.37 (83.7\%) \\
\hline
\cellcolor{gray!23}Memory utilization & 1.75 (100\%) & 2.50 (88.0\%) \\
\hline
\end{tabular}
\vspace{15pt}
\tightcaption{\name{}'s performance when metrics from system logs are selectively removed. Removed Marple, Jaeger, and cAdvisor features are shown in blue, red, and grey, respectively. Results list avg rank (\% in top-5) and are for Reddit.}
\label{t:log_features}
\vspace{0pt}
\end{table}

\para{System log analysis:}
To understand the relative importance of each metric in the system logs, we evaluated a variety of \name{} models that were trained with each log feature removed, in turn. Table~\ref{t:log_features} lists representative results. As shown, removing the per-switch \textit{packet counts} from the network logs led to the largest accuracy degradation, with a drop in average rank from 1.97 to 7.93 (for test\_generalize). Importantly, removing each considered feature led to marginal degradations in \name{}'s performance, highlighting their utility.

\section{Developer Study}
\label{s:devstudy}

To evaluate \name{}'s ability to accelerate end-to-end root cause diagnosis, we used our testbed (\S\ref{s:testbed}) to conduct a developer study. Developers were presented with the testbed's tools and logs, both with and without \name{}, and were tasked with diagnosing the root cause of multiple high-level user reports. In summary, developers with access to \name{} were able to correctly identify 90\% of the root causes (compared to 60\% without \name{}), and did so 72\% faster. 

\subsection{Study setup}

Our study involved 20 PhD students and postdoctoral researchers in systems and networking. All participants brought their own laptops, but debugging tasks were performed inside a provided VM for uniformity. Prior to the study, the authors delivered a 5-hour tutorial explaining the testbed and Sock Shop UI/code base; the study only involved Sock Shop to ease the developers' ability to become intimately familiar with the application to debug. For each tool (Marple, Jaeger, cAdvisor, tcpdump, \name{}), we described its logs, query language, and interface. Developers were given 1 hour to experiment with the testbed and resolve any questions.

During the study, developers were presented with a series of six debugging scenarios: 2 in-network faults for routing errors and congestion (targeting Marple), 2 system configuration faults for resource underprovisioning and component failures (targeting cAdvisor), and 2 application logic faults for branch condition and RPC errors (targeting Jaeger); we exclude end-host network faults due to time constraints. For each fault type, developers were randomly assigned to debug one fault using only the testbed's tools, and one also using \name{}. Ordering of the faults and tool assignments was randomized across participants to ensure a fair comparison.

For each fault, developers were presented with 1) a user report, 2) system logs for all testbed tools collected during the faulty run, and 3) the faulty testbed code. Developers were given 30 mins to diagnose each fault and provide a short qualitative description of the root cause. For example, a routing configuration error that disconnected Cassandra could be successfully reported as ``Cassandra could not receive any network packets, leading to missing page content.'' When a developer believed she had found the root cause, she informed the paper authors who verified its correctness. If incorrect, the developer was told to keep debugging until a correct diagnosis was generated, or 30 mins elapsed. Developers were unrestricted in their debugging methodologies, e.g., they were not required to use queries, though most did. Without \name{}, developers had to generate any query they wished to issue on their own; with \name{}, developers could generate queries or use the 5 suggested by \name{}.

\subsection{\name{}'s impact on root cause diagnosis}

\begin{figure}[t]
\centering
\includegraphics[width=.44\textwidth]{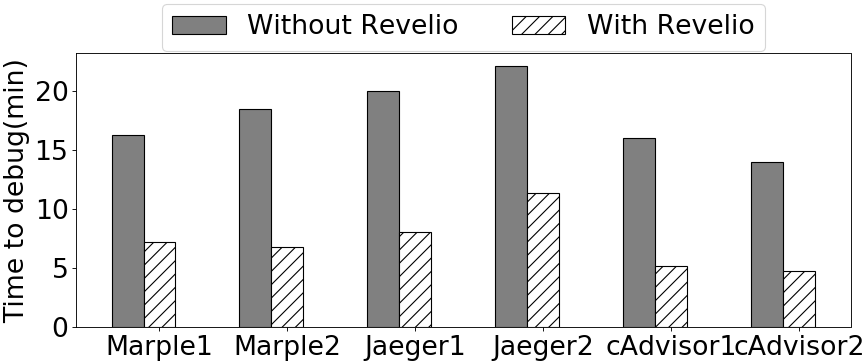}
\tightcaption{Summary of time saved in debugging each fault in our developer study. Bars represent average time spent across all developers who correctly identified the root cause.}

\label{fig:devsummary}
\end{figure}

The results of our developer study were promising and suggest that \name{} can be an effective addition to state-of-the-art debugging frameworks in terms of accelerating root cause diagnosis. Across all of the faults, \name{} increased the fraction of developers who could correctly diagnose the faults within the given time frame from 60\% to 90\%. Further, as shown in Figure~\ref{fig:devsummary}, \name{} sped up the average root cause diagnosis time by 72\% ($\sim$14 minutes) in cases where the developers were able to report the correct root cause.

After the study, we asked each developer qualitative questions about their experience with \name{}.

The most commonly reported benefit of \name{} was in shrinking the set of tools and queries that a developer had to consider. The primary gripe was with respect to \name{}'s UI, which is admittedly unpolished. Most importantly, the response to "Would you prefer to use existing systems and networking debugging tools with \name{}?", was ``yes'' for all 20 participants.

\section{Related Work}
\label{s:related}

We discuss the most closely related approaches here, and present additional related work (e.g., for triaging) in \S\ref{ss:related_more}.

\subsection{Debugging Tools for Distributed Systems}

There exist dozens of powerful data logging and querying tools for distributed systems~\cite{pivot,demi,jaeger,dapper,baggage,xtrace,canopy}, networks~\cite{netsight,pathqueries,marple,sdndebug,trumpet,cherrypick,pathdump,tcpdump}, and end-host stacks~\cite{gdb,dejavu,reverb,igor,whyline,lienhard08,dora,agrawal93,efficientslice,korel,polaris}. Although each tool provides powerful data logging and querying capabilities from different vantage points, two limitations exist. First, these tools are not coordinated and lack context about system-wide debugging. Thus, the cognitive burden of deciding which tools to use, when, and how (e.g., parameters) falls on developers. \name{} interoperates with these tools and alleviates this burden by automatically predicting helpful debugging queries.  Second, these tools ignore natural language inputs, including user reports characterizing {\em external behavior}. Our results (\S\ref{s:eval}) and prior work~\cite{netsieve,evolveordie} show that these inputs can be a rich source of debugging insights.

\subsection{Leveraging Natural Language Data Sources}

\para{Program debugging:}
NetSieve~\cite{netsieve} uses NLP to parse network tickets by generating a list of keywords and using a domain-specific ontology model to extract ticket summaries from those keywords; summaries highlight potential problems and fixes. While NetSieve automates parsing, much manual effort is still required in (1) offline construction of an ontology model, and (2) determining what constitutes a keyword. In contrast, \name{}'s models learn automatically from data, with minimal manual effort, and generate queries for root cause diagnosis rather than potential fixes from a restricted set of actions. Net2Text~\cite{net2text} translates English queries into SQL queries, issues those queries, summarizes the results, and translates them back into natural language for easy interpretation. \name{}, on the other hand, ingests high-level user issues and system logs; the unstructured and abstract nature of this input makes \name{}'s problem harder than Net2Text's.

\para{Program analysis and synthesis:}
NLP techniques have been utilized in multiple aspects of software development~\cite{ernstnlp}. Examples include detecting operations with incompatible variable types~\cite{ayudante} and converting natural language comments into assertions~\cite{toradocu}. More recently, NLP has also been used in code generation by converting developer-specified requirements in natural language to structured output in the forms of regular expressions~\cite{karthikregexp}, Bash programs~\cite{nl2bash}, API sequences~\cite{gu2016deep}, and queries in domain specific languages~\cite{nlpdsl}. Though these projects show the potential to extract meaning from natural language debugging data, they are limited to ingesting a single stream of data from a single subsystem. In contrast, \name{} combines and extracts meaning from varied input forms to construct structured debugging queries.

\section{Conclusion}
\label{s:conclusion}

\name{} employs ML to generate debugging queries from system logs and user reports to help developers find a problem's root cause faster. 
Much work remains before this general vision of an ML-enhanced debugging assistant for distributed systems is ready for production use. Notably, \name{} must present a uniform interface to all debugging tools (Figure 1) and learn from logs, reports, and queries in online fashion. Further, \name{} needs to be deployed in a live environment and evaluated with real faults.

Despite this, \name{} makes significant progress towards live deployment.

In particular, we learned: %

(1) a (modularized) pipeline of simple ML models is preferable to a single monolithic and complex model, (2) state-of-the-art NLP techniques such as BERT~\cite{bert} trained on news corpora can discern useful patterns from user reports, (3) GCNs serve as a universal translation layer to convert diverse query formats (e.g., Marple, Jaeger) into a common vector representation, (4) a uniform use of vector representations leads to an extensible architecture that allows easily 
incorporating new and diverse data sources, and (5) abstraction by ranking improves generalizability and scaling.

\label{lastpage}
\balance

\small{

\bibliographystyle{ACM-Reference-Format}
\bibliography{main}
}
\clearpage
\sloppypar
\appendix
\newpage
\section{Appendix}
\label{s:appendix}

\subsection{Additional Testbed Details}
\label{ss:testbed_more}

\subsubsection{Overview of Applications}
\label{ss:apps}

\para{Reddit:} Reddit is a popular discussion website whose three-tier backend architecture is representative of many distributed applications that utilize the monolithic architectural paradigm. In the front-end tier, HAProxy~\cite{haproxy} load balances traffic across web servers. The application tier, implemented using the Pylons framework for Python~\cite{pylons}, embeds the core application program logic and accesses data objects from the storage tier. The storage tier consists of three data stores: PostgreSQL~\cite{postgresql} is mainly used as a key/value store for objects such as accounts and comments; Cassandra~\cite{cassandra} is used as a key/value store for precomputed objects such as comment trees; and Memcache~\cite{memcached} is used for caching throughout the system. Reddit also uses the RabbitMQ message broker~\cite{rabbitmq} to manage asynchronous writes to the storage layer.

\para{Sock Shop:} Developed by Weaveworks, Sock Shop is an e-commerce application that employs a microservice-based backend architecture. Sock Shop incorporates 14 different microservices, including a user-facing Node.js frontend microservice, a shopping cart management microservice, a catalog microservice, and so on. Each microservice includes an application server whose logic is implemented in one of a variety of programming languages (e.g., Java, Go), and select microservices additionally operate an individually-managed datastore. For instance, separate MongoDB database instances~\cite{mongodb} are used for cart information, processed order transactions, and user profiles, while catalog information is stored in a MySQL database~\cite{mysql}. As with Reddit, RabbitMQ manages inter-microservice communication.

\para{Online Boutique:} Online Boutique is another microservice-based e-commerce platform from Google that includes 10 distinct microservices that are implemented in Python, Go, C\#, Java, and JavaScript. Microservices include a frontend HTTP server (implemented in Go), a payment microservice, a cart microservice, and an ad microservice. Each microservice operates its own datastore, e.g., the cart microservice stores a user's to-be-purchased items in Redis~\cite{redis}. Microservices communicate using the gRPC framework~\cite{grpc}.

\subsubsection{Overview of Debugging Tools}
\label{ss:tools}

\para{Marple:} Marple is a performance query language for network monitoring that uses SQL-like constructs (e.g., groupby, filter). To operate, Marple assigns each network switch and packet a unique ID, and supports queries that track 1) per-packet and per-switch queuing delays, and 2) user-defined aggregation functions across packets. In our implementation, switches log queue depths that each arriving packet encounters, and the packet's 5-tuple (src/dest ip addresses, src/dest ports, and protocol). This is sufficient to track queueing information and high-level statistics such as packet counts. We write queries in Marple to capture and track these values, and then use Marple's compiler to generate P4 programs~\cite{p4_behavioral} that can run directly on our emulated switches. Switches stream query results to a data collection server running on the same host machine for further analysis.

\para{tcpdump:} tcpdump is an end-host network stack inspector which analyzes all incoming and outgoing packets across all of the host's network interfaces. tcpdump's command line interface supports querying in the form of packet content filtering (e.g., by hostname, packet type, checksum, etc.), which can be applied at runtime or offline. In our implementation, tcpdump is configured to collect all network packet information in a pcap file, and filters are applied offline.

\para{Jaeger:} Uber's Jaeger framework is an end-to-end distributed systems tracing system which, like its predecessors Dapper~\cite{dapper} and Zipkin~\cite{zipkin}, implements distributed tracing according to the OpenTracing specification~\cite{opentracing}. With Jaeger, developers embed tracepoints directly into their system source code (or RPC monitoring proxies~\cite{envoy}) and specify custom state (e.g., variable values) to log at each one.

By aggregating tracepoint and timing information, Jaeger provides distributed context propagation so developers can understand how data values and control state flows across time and subsystems. We modified each application's source code (application tier for Reddit, and each microservice frontent for Sock Shop and Online Boutique) to include tracepoints for each function accessed during HTTP response generation. As per the examples provided by OpenTracing~\cite{opentracing}, at each tracepoint, we log the accessed variables, function execution duration, and any thrown exceptions. During execution, all tracepoint information is sent to a Jaeger aggregation server running on the same host machine for subsequent querying.

\para{cAdvisor:} Google's cAdvisor framework profiles the resource utilization of individual containers. To do so, cAdvisor runs in a dedicated container, which coordinates with a Docker daemon running on the same machine to get a listing of all active containers to profile (and the process ids that each owns). With this information, caAvisor uses the Linux cgroups kernel feature to extract resource utilization information for each container. We use cAdvisor's default configuration, in which the following values are reported every 1 second: instantaneous CPU usage, memory usage, and disk read/write throughput, and cumulative number of page faults. Resource usage information collected by cAdvisor is dynamically sent to a custom logging server running on the same host machine for subsequent querying.

\subsection{Additional Related Work}
\label{ss:related_more}
DeepCT~\cite{deepct} uses a GRU (Gated Recurrent Unit) and attention-based model to leverage discussions between  team members to accurately triage an incident. It incrementally learns knowledge from the discussions to better triage the incident. DeepTriage~\cite{deeptriage} leverages both textual and contextual data from incident reports. The textual data contains information like the incident title, summary and the initial discussion entries and the contextual data includes information about the service, severity of the incident, etc. The authors also identify that only few discussion items correspond to triaging and several of the subsequent discussion is on root cause analysis and logging troubleshooting steps which indicates that identifying the root cause requires several manual steps and is time consuming. While triaging identifies the team to detect root cause, it doesn't provide any hints as to the specific metrics and their relation to the subsystems to help developers.

DISTALYZER~\cite{distalyzer} leverages logs printed by distributed systems to identify the strongest association between performance and specific components. These log snippets are returned to the developer to identify the root cause of the incident. Due to the presence of several components in a distributed system and a bug can manifest itself in logs across different components, it is hard to find the most useful log by association alone and developer has to still sieve through all the logs.

\clearpage
\onecolumn
\subsection{Diagrams Illustrating Model Operation/Insights}

\begin{figure}[!h]
\centering
\includegraphics[width=\textwidth]{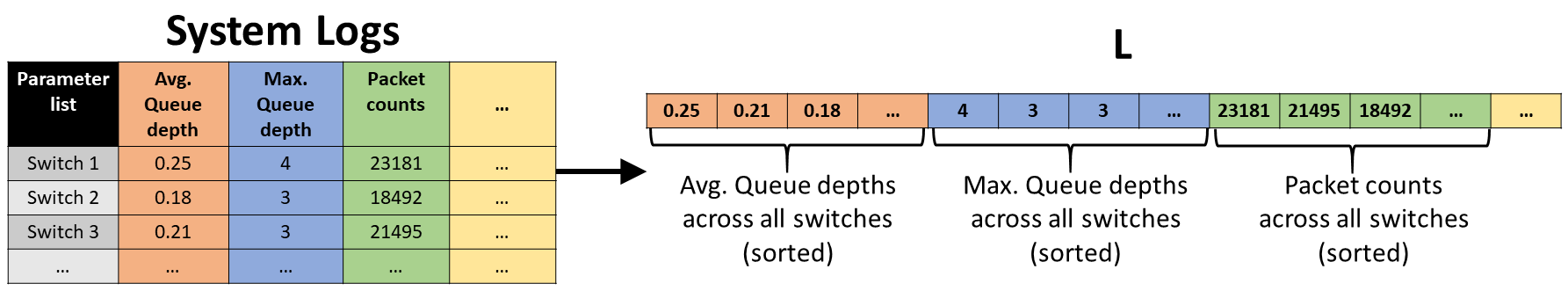}
\vspace{3mm}
\tightcaption{Example illustrating the generation of system log vector $L$; for simplicity, the example considers only network logs. Values for each feature (across switches) are first rank ordered, and then the resulting lists are concatenated to form $L$.}
\label{fig:feature_sorting}
\end{figure}
\vspace{10mm}

\begin{figure}[!h]
\centering
\vspace{1mm}
\includegraphics[width=0.9\textwidth]{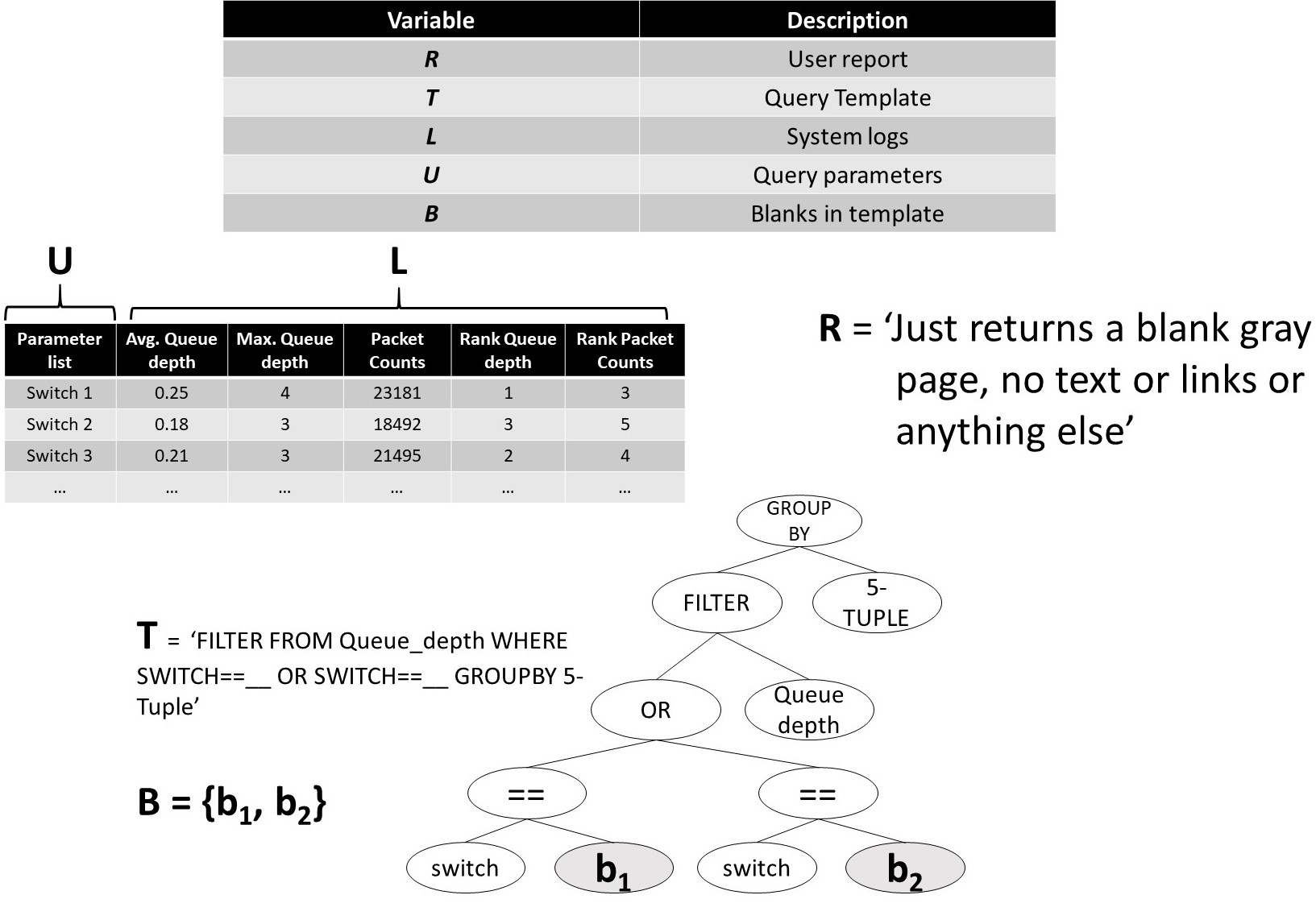}
\tightcaption{Example inputs for each input variable in \name{}'s model (variables are listed in Table~\ref{tab:revelio_ml} and at the top of this figure). This example is for a network (Marple) query. For the query template (T), the entire tree represents the template, while the parameters to be filled in are shaded in grey.}
\label{fig:revelio_inputs}
\end{figure}

\clearpage

\subsection{Diagrams for Testbed and Dataset}

\begin{figure}[!h]
\centering
\vspace{1mm}
\includegraphics[width=0.95\textwidth]{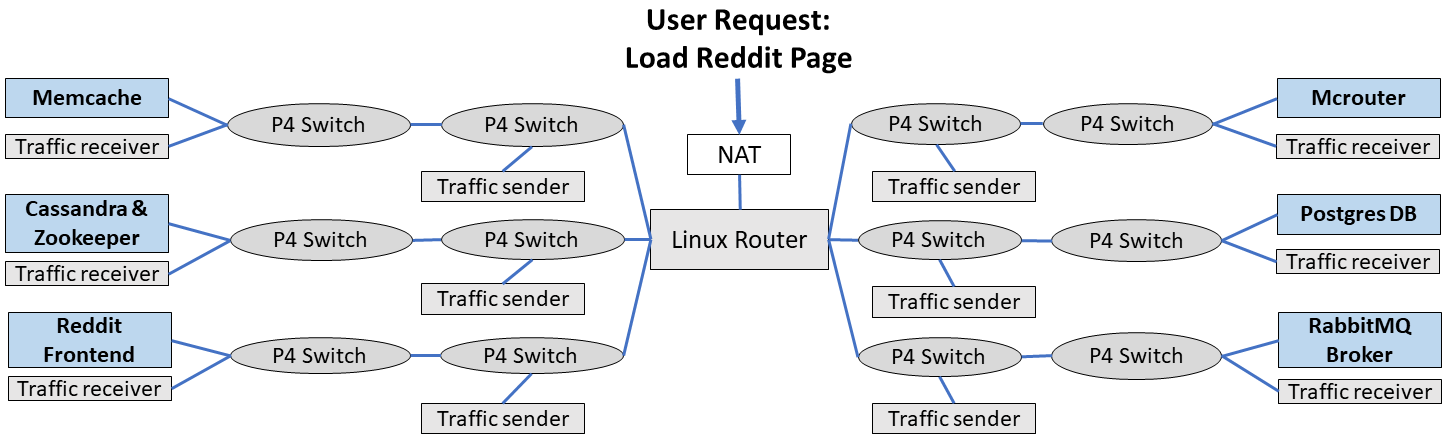}
\vspace{5mm}
\tightcaption{The topology of our distributed systems testbed for Reddit~\cite{reddit}. Each P4 switch has a congestion traffic sender/receiver to emulate different network conditions, and the testbed incorporates four recent debugging tools and a fault injection service. We illustrate the Sock Shop~\cite{sockshop} topology in Figure~\ref{fig:testbed}, and note that Online Boutique~\cite{hipstershop} follows the same architectural patterns.}
\label{fig:reddit_topology}
\end{figure}

\vspace{10mm}

\begin{figure}[!h]
\centering
\includegraphics[width=0.95\textwidth]{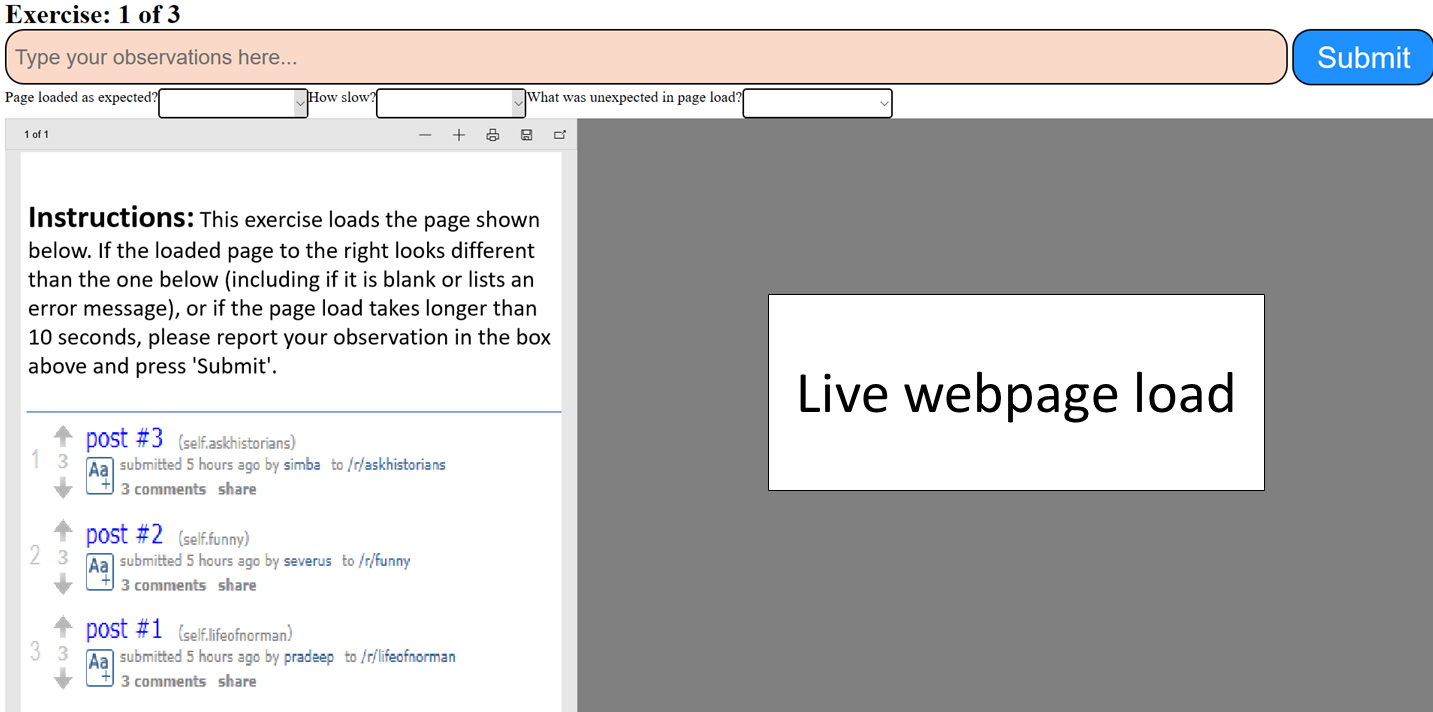}
\vspace{5mm}
\tightcaption{Screenshot of the UI presented to Mechanical Turk participants for inputting user reports. Instructions for the experiment are presented along with the live webpage load from the applications hosted in our distributed systems testbed. Users also choose from multiple choice options to report different aspects of the page load.}
\label{fig:ui}
\end{figure}

\clearpage

\begin{table*}[!h]
\small
\centering
\vspace{8mm}
\begin{tabular}{|c|}
\hline
\textbf{User report text} \\
\hline
there is nothing on the page, it is empty, nothing to click on\\
\hline
`you broke reddit' with the cartoon showed up\\
\hline
Page took forever to load. Sat at the gray screen for almost a minute, entirely too long..\\
\hline
I clicked to expand for comments, and page went away and defaulted to a grey screen.. No Page\\
\hline
`Funny 500 Page Message 8' message below that. Blank otherwise. . Page does not include any usernames\\
\hline
\end{tabular}
\vspace{4mm}
\tightcaption{Examples of text in user reports collected from Mechanical Turk participants.}
\label{t:report_examples}
\end{table*}

\begin{table*}[!h]
\small
\centering
\vspace{8mm}
\begin{tabular}{ |c | c| c|}
\hline
\textbf{Fault Type} & \textbf{Number of Faults} & \textbf{Example} \\
\hline
Resource underprovisioning & 15 & Reducting the CPU quota\\
& & for the docker container running PostgreSQL\\
\hline
Component failures & 15 & Take down container for a given microservice \\
\hline
Subsystem misconfigurations & 12 & Incorrectly configure hostname of a database\\
\hline
Network congestion & 13 & Generate significant network \\
& & cross-traffic between hosts for different microservice\\
\hline
Network-level misconfigurations & 16 & Incorrect firewall rules at routers to drop\\
& & or forward packets on an incorrect interface\\
\hline
Subsystem/Source-code bugs & 16 & Negated if condition resulting in different execution path\\
\hline
Incorrect data exchange & 15 & Alter function signature within \\
& & a microservice, triggering argument violations\\
\hline
\end{tabular}
\vspace{4mm}
\tightcaption{Overview of faults injected into our distributed systems testbed. Numbers listed are for Sock Shop\cite{sockshop}.}
\label{t:fault_list}
\end{table*}

\end{sloppypar}

\end{document}